# Evidence of ferromagnetic short-range correlations in cubic La$_{1-x}$Sr$_x$MnO$_{3-\delta}$ (x=0.80, 0.85) above antiferromagnetic ordering

Aga Shahee*, Kiran Singh, R. J. Choudhary and N. P. Lalla

*UGC-DAE Consortium for Scientific Research, University campus, Khandwa road Indore, India- 452001*

We report the existence of ferromagnetic correlations (FMC) in paramagnetic (PM) matrix of cubic La$_{1-x}$Sr$_x$MnO$_{3-\delta}$ ($x$ = 0.80, 0.85) well above its coupled structural, magnetic and electronic phase transitions. The dc-magnetization vs temperature [$M(T)$] behaviour under different magnetic fields (from 100 Oe to 70 kOe) shows the presence of short range magnetic correlations up to ($T_{FMC}$ ~) 365 K, far above the antiferromagnetic ordering temperatures ($T_N$ =) 260 K and 238 K for $x$=0.80 and 0.85, respectively. More importantly the observed short-range FMC survive even up to 70 kOe, which indicates their robust nature. The temperature region between $T_N$ to $T_{FMC}$ is dominated by the presence of correlated ferromagnetic (FM) entities within the PM matrix and stabilized due to A-site chemical disorder. Our results further illustrate that for the studied compositions, the oxygen off-stoichiometry does not have any significant effect on the nature and strength of these FM entities; however, FM interactions increase in the oxygen deficient samples. These compositions are the unique examples, where the presence of FMC is observed in an undistorted basic cubic perovskite lattice well above $T_N$ and therefore are novel to understand the physics behind the colossal magneto-resistance effect.

**1 Introduction** Manganites offer rich and complex magnetic, electronic and structural phase diagram[1,2]. This richness is due to the fact that even a slight change in the composition, preparation conditions and/or external conditions can cast drastic changes in their physical and structural properties[1,2]. The origin of complexity lies in the competing ordering tendencies of different degrees of freedoms i.e. charge, spin, orbital and lattice, which produce nearly degenerate ground states with significantly different physical properties[1-3]. It is this competition, which is responsible for the phenomenon of phase separation[2] in manganites.

Among the various types of inhomogeneous phases, the existence of ferromagnetic (FM) clusters in paramagnetic (PM) matrix, well above its FM Curie temperature ($T_C$), is a established and important phenomenon[2,4]. Such scenario of short-range FM correlations (FMC) was first predicted by Griffiths[5] for randomly diluted Ising-ferromagnets. Later on Bray and Moore[6,7] generalized the Griffith's idea to any disordered magnetic system containing bond distributions, and termed this as Griffiths phase (GP). Thereafter, such GP-like behaviour has been reported in various manganites[8-18] as well as other compounds[19-21]. The observation of GP-like behaviour in manganites is mostly attributed to the structural disorder[8,9,13-15], strain field[10,18], phase inhomogeneity[11], acoustic spin wave stiffness[15] and oxygen vacancies[17]. In 2008, Souza et al.[22] suggested that the GP-like treatment to a short range FMC in disorder manganites is inappropriate and is due to FM-polaron[23]. Such FM polaron formations are due to intrinsic chemical and structural disorder. This was further supported by Rozenberg et al.[24-27]. Turcaud et al.[28] related such behaviour above $T_C$ with the transition temperature of the low-temperature (high magnetic field) magnetic phase. Thus the descriptions of Souza et al.[22] and Turcaud et al.[28] are fundamentally different from the GP description. Here after, we will call such behavior as FMC. Some experimental reports[8,22,29,30] claim that FMC plays a key role behind the physics of colossal magnetoresistance (CMR)[1,2], while Jiang et al.[15] contradict this.

Thus, whether we discuss GP-singularity or FM polarons, both demand the presence of some degree of chemical and/or structural disorder. In manganites the sources of such disorder are the A-site substitutional (chemical) disorder and the randomly distributed $Mn^{3+}$ ions, which may cause random JT-distortion of $Mn^{3+}O_6$ octahedra. To the best of our knowledge, the manganites showing presence of FMC in PM matrix not only have the above mentioned disorders but also possess coherent structural distortion of their basic perovskite lattice[8-21]. In perovskite manganites, the coherent structural distortion is a combine effect of cooperative JT-distortions and octahedral tilt[31]. Such octahedral tilts cause the bending of Mn-O-Mn bonds, resulting in the reduction of exchange hopping amplitude. A straight Mn-O-Mn bonds (angle=180$^o$), which are present in the basic cubic perovskite manganites, favors strongest FM interaction. Variation in the indirect exchange hopping amplitude will have significantly different implications on the above discussed disorder induced FMC in PM matrix in manganites i.e. the behaviour of FMC in cubic perovskite manganites may be fundamentally different from that of the distorted perovskite manganites. As referred above, there are several studies[8-18] in distorted perovskite manganites but none on cubic perovskite manganites. Thus the possible presence of FMC in PM matrix of a disordered cubic perovskite manganite phase appears to be worth exploring.

Keeping in view the above motivation, we have explored the occurrence of FMC in a cubic perovskite system $La_{1-x}Sr_xMnO_{3-\delta}$ ($x$ = 0.80 and 0.85). These samples possess interesting structural and physical properties as investigated by Fujishiro et al.[32], Hemberger et al.[33] and Chmaissem et al.[34]. These samples undergo structural phase transition from cubic (*Pm-3m*) to tetragonal (*I4/mcm*) concomitant with C-type antiferromagnetic (AFM) spin ordering and a jump in resistivity at ~ 260 K and ~ 240 K for $x$ = 0.80, 0.85, respectively[33,34]. Thereafter, Bindu et al.[35,36] reported the existence of charge ordering ($T$ < 265 K) and structural phase coexistence in $La_{1-x}Sr_xMnO_{3-\delta}$ ($x$ = 0.80). In all these previous reports the magnetic studies have been done maximum up to 400 K[32-34] at single $H$ with prime interest to know the AFM ordering temperature ($T_N$). Recently, we have studied the effect of oxygen vacancies in these compounds and observed that the oxygen vacancies completely suppress the coupled structural, magnetic and electronic phase transitions[37,38]. We have observed that oxygen vacancies actually stabilize the above reported charge ordered phase in a cubic perovskite lattice[37]. But till date there is no detailed investigations of magnetic behaviors of these materials particularly to explore their magnetic behavior above $T_N$. Here, we report high $T$ magnetic studies at different $H$ between 100 Oe-70 kOe to demonstrate the existence of strongest FMC in a cubic perovskite lattice of $La_{1-x}Sr_xMnO_{3-\delta}$ ($x$ = 0.80, 0.85), which sustains even at $H$ as high as 70 kOe.

**2 Experimental Details** Single phase polycrystalline samples of $La_{1-x}Sr_xMnO_{3-\delta}$ (with $x$ = 0.80 with δ = 0.01, 0.12 and 0.85 with δ = 0.02, 0.14) were synthesized by solid-state reaction method followed by argon annealing for creating oxygen off-stoichiometry. The structural phase purity and chemical homogeneity of the samples was characterized using x-ray diffraction and EDAX analysis in transmission electron microscopy (TEM). For estimating oxygen off-stoichiometry, iodometric titration was carried out. The details of synthesis, characterization, and low-$T$ magnetic and structural studies are reported elsewhere[37,38]. The main goal of the present study is to explore the magnetic properties above $T_N$. The dc magnetization measurements were performed at different $H$ between 100 Oe to 70 kOe in zero field cooled (ZFC) and field cooled (FC) modes using a SQUID-VSM magnetometer (QD-SVSM) in temperature $(T)$ range of 50 - 600 K. Isothermal magnetization measurements $M(H)$ were performed at different $T$ with field ranging from +70 kOe to -70 kOe. For measurements beyond 325 K, oven option of QD-SVSM was used.

**3 Results and Discussion** Fig. 1 (a-d) represents the low-field (100 Oe) and high-field (50, 70 kOe) $M(T)$ behavior in ZFC and FC modes. It shows a clear cusp around $T_N \sim 238$ K and $\sim 260$ K for oxygen stoichiometric $La_{0.15}Sr_{0.85}MnO_{2.98}$ and $La_{0.20}Sr_{0.80}MnO_{2.99}$ samples, respectively, associated to the onset of long range C-type canted AFM spin ordering [33,34] concomitant with cubic to tetragonal structural transitions [37,38]. In the case of $La_{0.15}Sr_{0.85}MnO_{2.98}$, in addition to $T_N$, there is another anomaly around 100 K, whose origin is still unclear. Similar anomaly is also observed by others [33,34]. Our magnetic results are fully consistent with earlier reports [34] confirming the good quality of our samples. Oxygen deficient samples do not show any type of long range ordering (structural/magnetic) as already reported [37,38].

The most interesting feature of these four samples is the anomaly emerging in $M(T)$ data at ~ 365 K, see Fig. 1. The study of this anomaly is the main goal of this article. Although this anomaly in the $M(T)$ data appear to mimic a typical PM to FM like transition, but it is not the case here. Similar feature was also observed by Fujishiro et al.[32] for $x = 0.82$ and 0.90 but they did not pay much attention to this. To explore this anomaly in detail, ZFC and FC $M(T)$ was performed under different $H$ up to 600 K (for $x = 0.80$). The sharp increase of magnetization at ~ 365 K is accompanied by significant divergence in the ZFC and FC $M(T)$ curves at low $H$. The bifurcation in FC and ZFC curves above $T_N$, indicates the presence of FMC. As compared to nearly oxygen stoichiometric samples, the ZFC and FC bifurcation decreases in oxygen deficient samples, see Fig. 1 (b) and (d). The sharp anomaly in $M(T)$ becomes broader at higher $H$. This will be discussed in detail in the following.

Fig. 2 shows $M(H)$ results at different $T$ below and above $T_N$. At lower $T$, $M(H)$ loops show small FM like hysteresis, but no saturation even at 70 kOe. This indicates canted AFM like behaviour. For all samples the $M(H)$ curves at ~365 K (i.e. $T > T_N$ of stoichiometric samples), show a non-linear behaviour at low $H$ (<10 kOe). This behaviour at low $H$ indicates the presence of weak magnetization, either due to the presence of short or long range FMC. In order to investigate, whether FMC are short or long range, $M^2$ was plotted against $H/M$ (i.e. an Arrott plot) at 300 K for all the studied compositions as shown in Fig. 3(a-d). It is well established that if the linear extrapolation of the high-field portion of the Arrott plot yields a positive intercept on $M^2$-axis, it indicates existence of spontaneous magnetization with long-range FM order, while a negative intercept indicates short-range FM correlations[10,11]. In our studies the above mentioned linear extrapolation yields a negative intercept on $M^2$-axis for all the studied compositions, see Fig. 3. This shows the presence of short-range FMC in these samples.

The low field (100 Oe) $\chi^{-1}$ vs $T$ plots for $x = 0.80$ and $0.85$ are presented in Fig. 4(a) and 4(b) respectively. The Curie-Weiss (CW) fitting T ranges are 600 K to 460 K for $x=0.80$ and 470 K to 405 K for $x= 0.85$. We observe a deviation from the CW law below ~ 460, 405 K (here after referred as $T^*$) for $x = 0.80$ and 0.85 respectively. This indicates that the region between $T^*$ and $T_N$ is not pure PM but have the existence of magnetic polaron/FMC. The continuous decrease in $\chi^{-1}$ below $T^*$ during cooling indicates continuous increasing concentration of the FM polarons. These continuously increasing magnetic polarons start interacting with each other below certain temperature say $T_{FMC}$ and give rise to sudden increase in $M(T)$. The $T_{FMC}$ has been determined from the peak value of $d\chi^{-1}/dT$ vs $T$, see inserts of Fig. 4. Its value corresponds to ~ 364 K, 367 K, 365 K and 360 K for $La_{0.20}Sr_{0.80}MnO_{2.99}$, $La_{0.20}Sr_{0.80}MnO_{2.88}$, $La_{0.15}Sr_{0.85}MnO_{2.98}$ and $La_{0.15}Sr_{0.85}MnO_{2.86}$, respectively. This has commonly been recognized as the signature of the "Griffith's- like" singularity[8-21]. As the polaron formation has direct impact on the magnetization of a sample, depending on whether the spin correlation is FM or AFM, even in the PM region it will either enhance or suppress the effective magnetic momentum ($\mu_{eff}$). Therefore in order to probe the nature of polarons we have determined $\mu_{eff}$ of the studied samples using CW fit. Table 1 shows the experimentally determined and theoretical calculated spin only $\mu_{eff}$ values. It can be noted that the experimental $\mu_{eff}$ values of studied samples deviate from the corresponding expected theoretical values, well beyond the experimental error from the expected theoretical values. This indicates that the present observation may be a case of magnetic polaron formation. An experimental $\mu_{eff}$ more than the theoretical $\mu_{eff}$ corresponds to magnetic polarons with FM interaction and vis a vis. Further the $\chi^{-1}(T)$ below $T_{FMC}$ shows a sharp down turn, which is considered to be the hallmark of stabilization of interacting FM polarons in PM matrix[10-12,23,24] and rejects the possibility of smeared phase transition[39,40,41] which will show an upturn in $\chi^{-1}(T)$ curves. The oxygen off-stoichiometric samples show similar $\chi^{-1}(T)$ behaviour as that of stoichiometric ones with only a few kelvin shifts in $T_{FMC}$ accompanied with the weakening of AFM correlations. These observations illustrate that although oxygen vacancies completely destroy the coupled low-$T$ structural and magnetic phase transition[37,38] but it does not cast any major effect on $T^*$, $T_{FMC}$, and the strength of FMC.

In order to further confirm the FMC, $\chi^{-1}(T)$ was investigated at different $H$ for $x = 0.80$, see Fig. 5(a,b). The softening of the down turn ~ $T^*$ with increasing $H$ indicates non-analytic behavior of magnetization and thus further supports the existence of short range FMC. The resemblance of our experimental observations with other reports[8-28] confirm the existence of the FMC. The observed $T_{FMC}$ in the present case is found to coincide with the $T_C$ of $La_{0.6}Sr_{0.4}MnO_3$[33]. This $T_C$ is infact the highest of the $La_{1-x}Sr_xMnO_3$ series. Thus the observation further confirms that the observed behavior is associated with FMC [6,7].

From the above discussion, it is evident that the studied materials show the existence of short range FMC in PM matrix. Now, we will discuss its robustness in comparison with other such reports in manganites and other magnetic oxides. It is imperative to mention here that in the earlier reports, such FMC were totally masked by the application of only few hundred/thousand Oe of $H$[8-24], making $\chi^{-1}(T)$ almost straight like a CW behavior, i.e. the PM susceptibility overwhelms that of the FMC. Surprisingly in our case even after 70 kOe the $\chi^{-1}(T)$ in FMC region is non-linear i.e. the FMC remains unmasked. It infers that the FMC in these samples is quit robust and dominates the PM contribution even at 70 kOe. The robust nature of FMC in these samples is due to cubic symmetry of the basic perovskite lattice, which gives rise to straight Mn-O-Mn bond with equal bond lengths. These bond angles provide widest band width and such octahedral configuration in manganites energetically favors FM interaction resulting in the highest $T_C$ in La$_{1-x}$Sr$_x$MnO$_3$ phase diagram[33]. Thus the extra stability of FMC is due to cubic symmetry. It is reported that in the case of cubic SrMnO$_3$, 1-2% Ce doping leads to the formation of ferromagnetic polarons[42], which vanishes on further Ce doping due to the occurrence of tetragonal distortion associated with AFM ordering. Similarly, few angstrom size magnetic polaron has been observed in long range non JT-distorted La$_{1-x}$Sr$_x$MnO$_3$[43]. These magnetic polarons persist well above $T_C$ in the form of local JT-distortion. All these results indicate that "cubicity" of the basic perovskite provides extra stability to polaronic states leading to FMC. These polarons can occur either in glassy or dynamic state in order to keep the cubic symmetry intact as may be the case in our samples, where static polaron with FM dimmers has been mostly observed in orthorhombically distorted phases[44,45,46]. Thus the most robust FMC observed in our samples as compared to earlier reports on distorted perovskite manganites, indicates the strengthening of short range FMC by its favorable high symmetric cubic lattice. In our case the oxygen stoichiometric samples show magneto-structural phase transition[37-38] whereas non-stoichiometric ones remains cubic down to low temperature but their magnetization, $T_{FMC}$ and $T^*$ remains nearly the same, thus does not support Turcaud et al.[28] model for low temperature (high magnetic field) phase. In manganites, a broken magnetic Mn–O–Mn network in perovskite lattice is expected to cause drastic effect on FMC if it is caused by the GP-like singularity. But in the present case, the observed FMC is nearly unaffected for oxygen-deficient samples, which have broken Mn-O-Mn network. Thus the current observation does not support a GP-like singularity. Our results also directly indicate its origin in A-site quench chemical disorder, which favors the idea of Souza et al.[22]. Above that the cubic symmetry favors FM interaction making the FMC robust in the PM matrix below $T_{FMC}$. In the following we will briefly discuss the possible origin of FMC in the studied samples.

The single valence state of Mn ion in un-doped parent compounds SrMnO$_3$/LaMnO$_3$ lead to magneto-electronically homogeneous AFM interactions. Divalent doping in these systems makes them intrinsically inhomogeneous due to the random distribution of cations of different sizes and mix-valent spin states of Mn. Thus strong competition between different ordering tendencies starts to evolve, which leads to a complex electronic/magnetic phase diagram. Through transmission electron microscopy, the existence of nano-phase electronic and magnetic inhomogeneities in chemically homogeneous samples has been observed[47]. The disorder in doped manganites is inevitably introduced since they are a solid solution of different A-site cations[48]. In our samples, quenched A-site chemical disorder is present due to the random distributions of $Sr^{2+}$ and $La^{3+}$ ions, which induce A-site ionic radii mismatch variance ($s^2 = \sum x_i r_i^2 - <r_{avg}>^2$) of ~ 0.001094 and 0.000816 $Å^2$ for $x$=0.80 and 0.85, respectively. Such chemical disorder may give rise to $Sr^{2+}$ and $La^{3+}$ ions rich regions. The FMC behaviour could come from an inhomogeneous distribution of the perovskite A-site cations. Macroscopic $La^{3+}$ rich region may enhance ferromagnetic correlations in contrast to macroscopic $Sr^{2+}$ rich region which would enhance antiferromagnetic correlations. In $La_{1-x}Sr_xMnO_3$ phase diagram, the highest AFM and FM ordering temperature is reported for $SrMnO_3$ and disorder $La_{0.6}Sr_{0.4}MnO_3$ with $T_N$ ~ 280 K and $T_C$ ~ 370 K respectively[33]. This causes a nearly 110 K high FM ordering temperature as compared to AFM ordering temperature in $La_{1-x}Sr_xMnO_3$ series. Thus the low TN value of the Sr2+-rich AFM region as compared to the TC of La3+-rich FM region will stabilize the FMC in a PM matrix up to the TC of La3+-rich region. In addition a finite variance of the A-cation radii distribution has already been reported to induce FMC in PM matrix in manganites[19,22,24-27]. Further the composition of our samples lie between pure C-type AFM (for $x$ = 0.75) and pure G-type AFM (for $x$ = 1.0) spin states. This dilutes the effective magnetic interactions to promote a tendency towards frustration and complex spin states[49]. Thus our results suggest that the quenched A-site chemical disorder and competing FM/AFM interactions are responsible for the occurrence of FMC in PM matrix in manganites. The robustness of the observed FMC against a field as high as 70 kOe suggest the existence of large phase fraction of FMC and a strong ferromagnetic coupling between these FMC below $T_{FMC}$ in these samples.

**4 Conclusion** In conclusion, we have demonstrated the presence of strong FMC in PM matrix of cubic perovskite $La_{1-x}Sr_xMnO_{3-\delta}$ (with $\delta$ = 0.01, 0.12 for $x$ = 0.80 and $\delta$ = 0.02, 0.14 for $x$ = 0.85) by revealing a downturn of $\chi^{-1}(T)$ curves with $T^*$ ~ 460 and 405 K and $T_{FMC}$ ~ 365 K. The $T_{FMC}$ is much higher than $T_N$ for the stoichiometric samples. The existence FMC up to $T_{FMC}$, well above AFM ordering for stoichiometric samples, suggest that the magnetic transition follow a PM to short range magnetic ordering (at $T_{FMC}$) and then to long range AFM ordering (at $T_N$) rather than a direct PM to AFM transition as reported earlier. The occurrence of FMC is due to quenched A-site disorder and competing magnetic interactions. The existence of FMC in a cubic lattice indicates that structural distortion is not a prerequisite for its existence, rather A-site disorder and competing magnetic interactions are the primary parameters for its stabilization. These results have broader significance for further exploration of disorder induces FMC in other such undistorted perovskite systems. This will be further helpful in developing theoretical model for explaining how such robust FMC exists in a cubic lattice.

**Acknowledgements** Aga Shahee would like to acknowledge CSIR-India for financial support.

**Table Caption:-**

**Table 1.** Fitting parameters Curie constant (C) and PM Curie temperature ($\theta_P$) obtained from Curie-Weiss law. Experimentally determined and theoretically calculated spin only effective magnetic moment ($\mu_{eff}$) of Mn ions.

**Figure caption:-**

**Figure 1** Temperature dependence of ZFC and FC magnetization curves under two selective magnetic fields for (a) $La_{0.20}Sr_{0.80}MnO_{2.99}$, (b) $La_{0.20}Sr_{0.80}MnO_{2.88}$, (c) $La_{0.15}Sr_{0.85}MnO_{2.98}$ and (d) $La_{0.15}Sr_{0.85}MnO_{2.86}$.

**Figure 2** Isothermal magnetization vs magnetic field dependence of (a) $La_{0.20}Sr_{0.80}MnO_{2.99}$, (b) $La_{0.20}Sr_{0.80}MnO_{2.88}$, (c) $La_{0.15}Sr_{0.85}MnO_{2.98}$ and (d) $La_{0.15}Sr_{0.85}MnO_{2.86}$ at different *T*.

**Figure 3** Arrott plots at 300K for (a) $La_{0.20}Sr_{0.80}MnO_{2.99}$, (b) $La_{0.20}Sr_{0.80}MnO_{2.88}$, (c) $La_{0.15}Sr_{0.85}MnO_{2.98}$ and (d) $La_{0.15}Sr_{0.85}MnO_{2.86}$.

**Figure 4** Temperature dependence of the inverse dc magnetic susceptibility ($\chi^{-1}$) under 100 Oe for (a) $La_{0.20}Sr_{0.80}MnO_{3-\delta}$ ($\delta$=0.01, 0.12) and (b) $La_{0.15}Sr_{0.85}MnO_{3-\delta}$ ($\delta$=0.02, 0.14). The Curie-Weiss fits are represented by solid straight lines. T* represents the lowest temperature down to which $\chi^{-1}$ *vs T* follows a Curie-Weiss behaviour. The respective inserts present $d\chi^{-1}/dT$ *vs T* under 100 Oe magnetic field.

**Figure 5** Temperature dependence of the inverse dc magnetic susceptibility ($\chi^{-1}$) under different magnetic fields for $La_{0.20}Sr_{0.80}MnO_{3-\delta}$ (a) $\delta$=0.01, (b) $\delta$=0.12.

| Sample | $\mu_{eff}$ / fu Experimental | $\mu_{eff}$ / fu Theoretical | C Curie constant | $\theta_P$ PM Curie temperature |
|---|---|---|---|---|
| $La_{0.20}Sr_{0.80}MnO_{2.99}$ | 3.82±0.02 $\mu_B$ | 4.12 $\mu_B$ | 1.831±0.022 | -52±1 |
| $La_{0.15}Sr_{0.85}MnO_{2.98}$ | 3.70 ±0.02 $\mu_B$ | 4.09 $\mu_B$ | 1.714±0.023 | -8±1 |
| $La_{0.20}Sr_{0.80}MnO_{2.88}$ | 4.46 ±0.02 $\mu_B$ | 4.35 $\mu_B$ | 2.491±0.021 | -15±1 |
| $La_{0.15}Sr_{0.85}MnO_{2.86}$ | 4.74±0.02 $\mu_B$ | 4.34 $\mu_B$ | 2.821±0.023 | 37±1 |

Table 1.

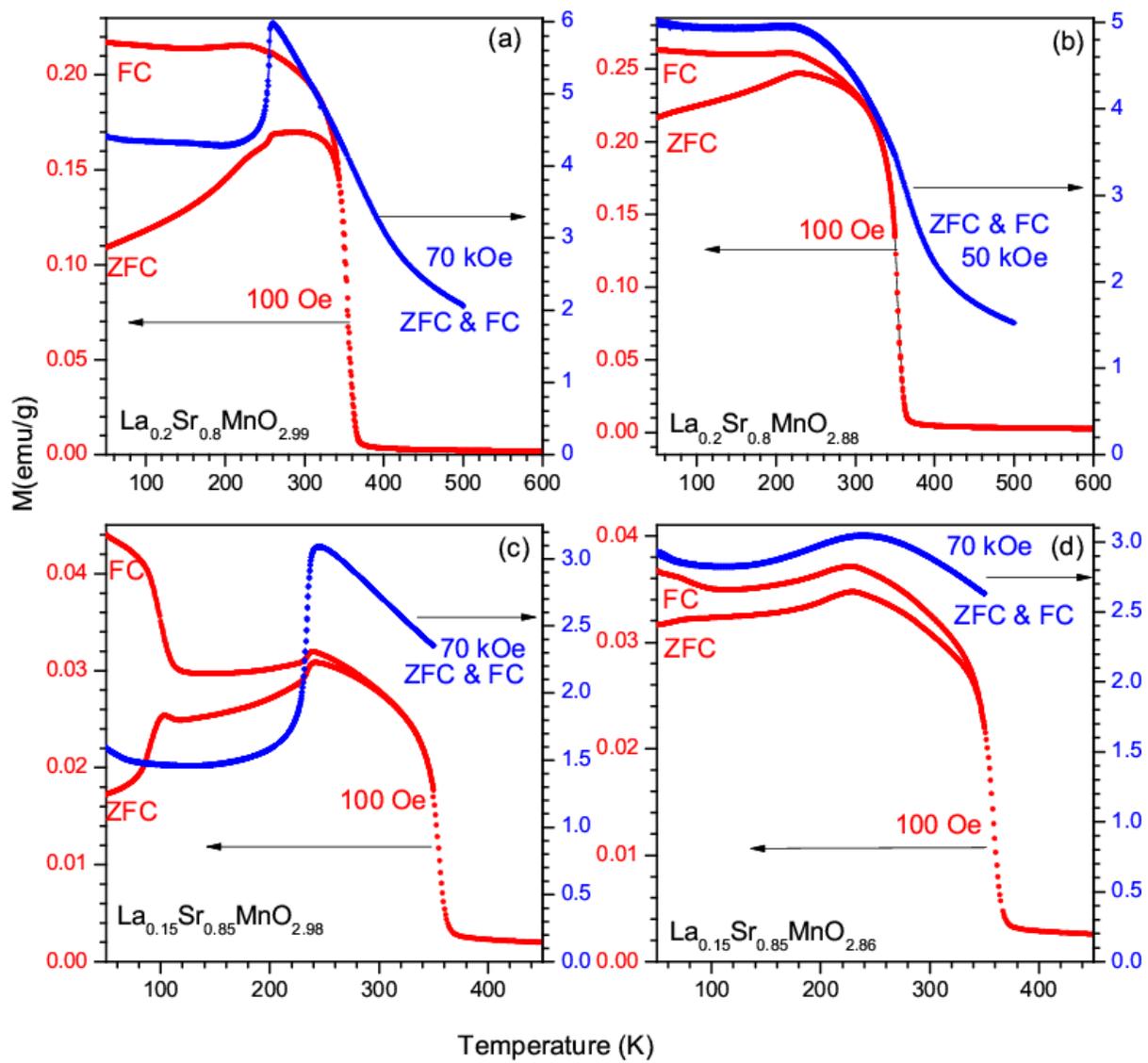

FIG 1.

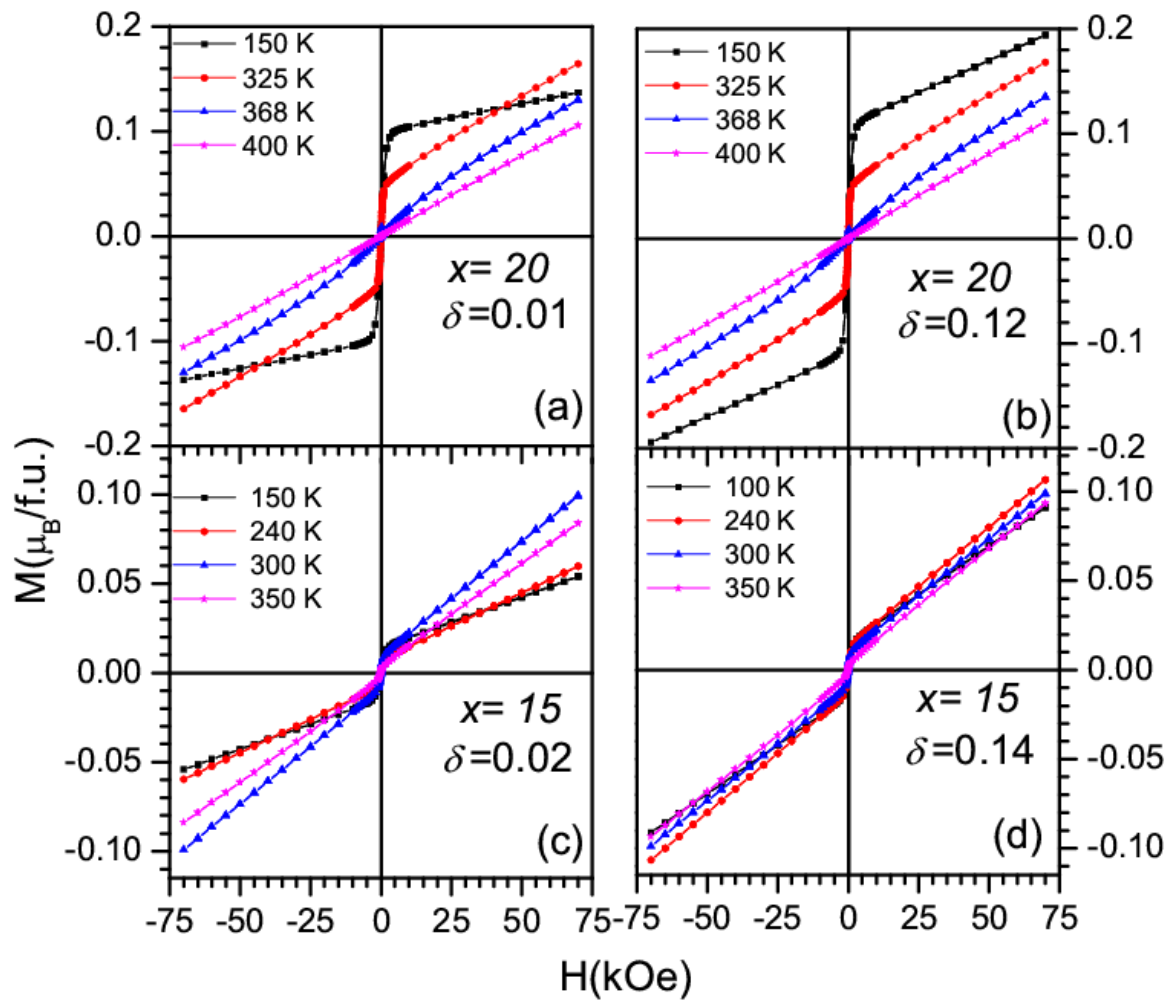

FIG 2.

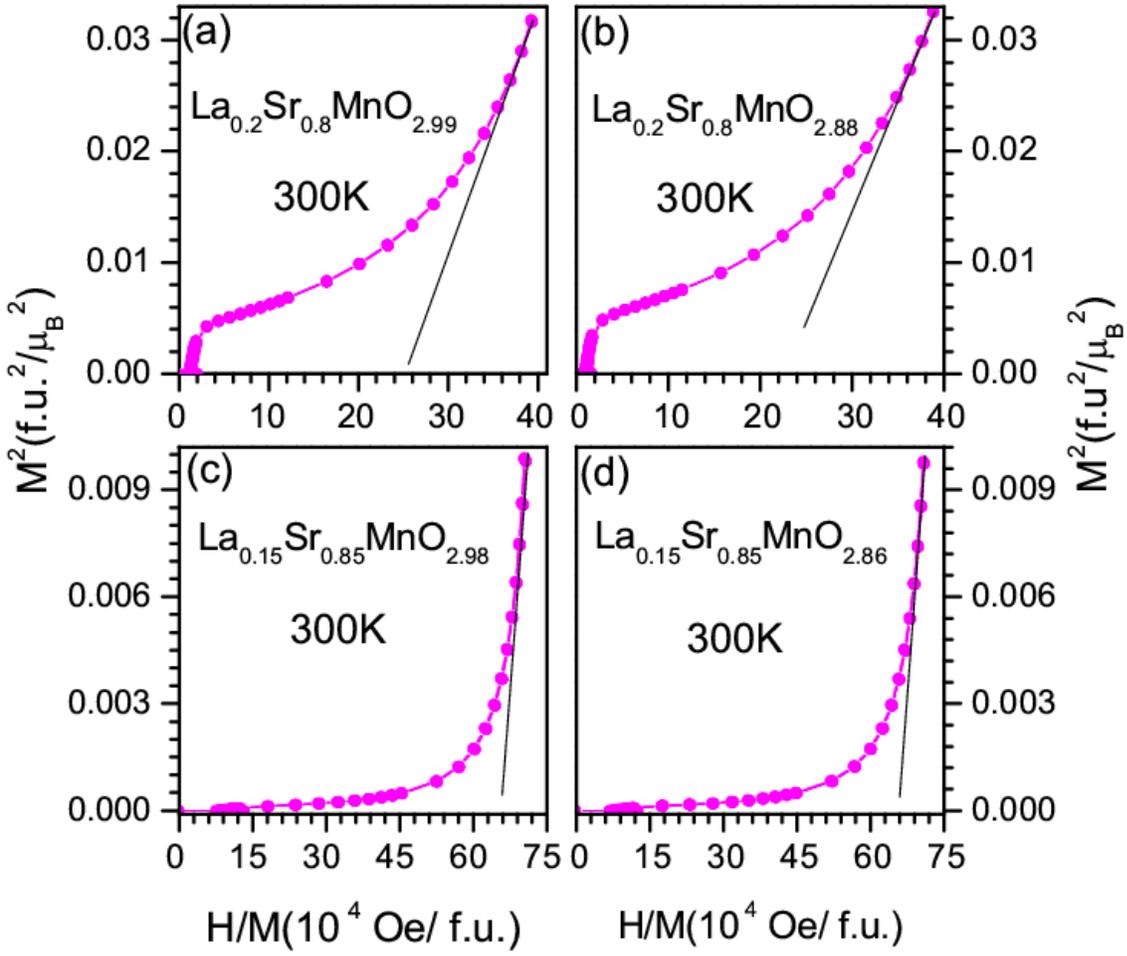

FIG 3.

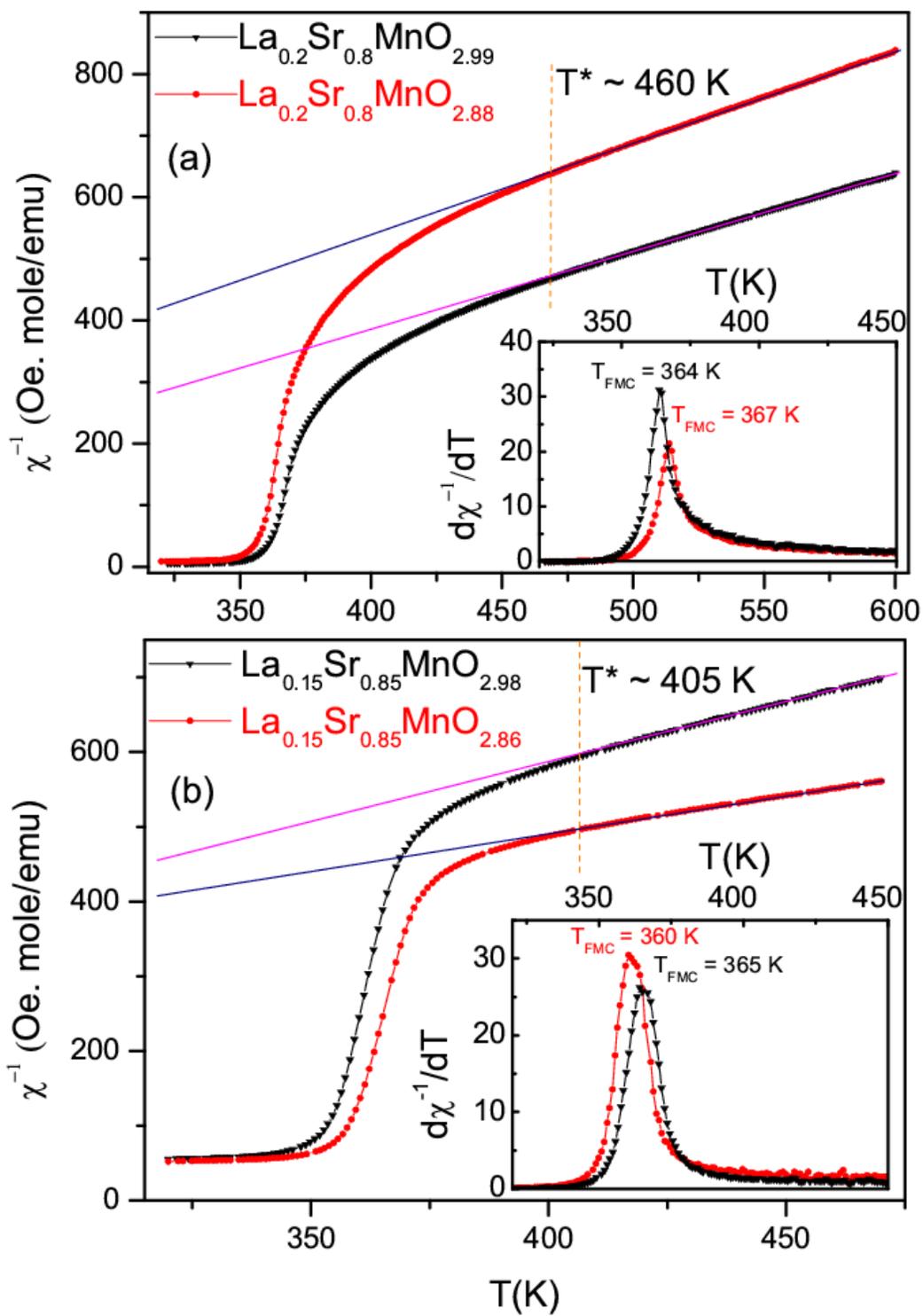

FIG 4.

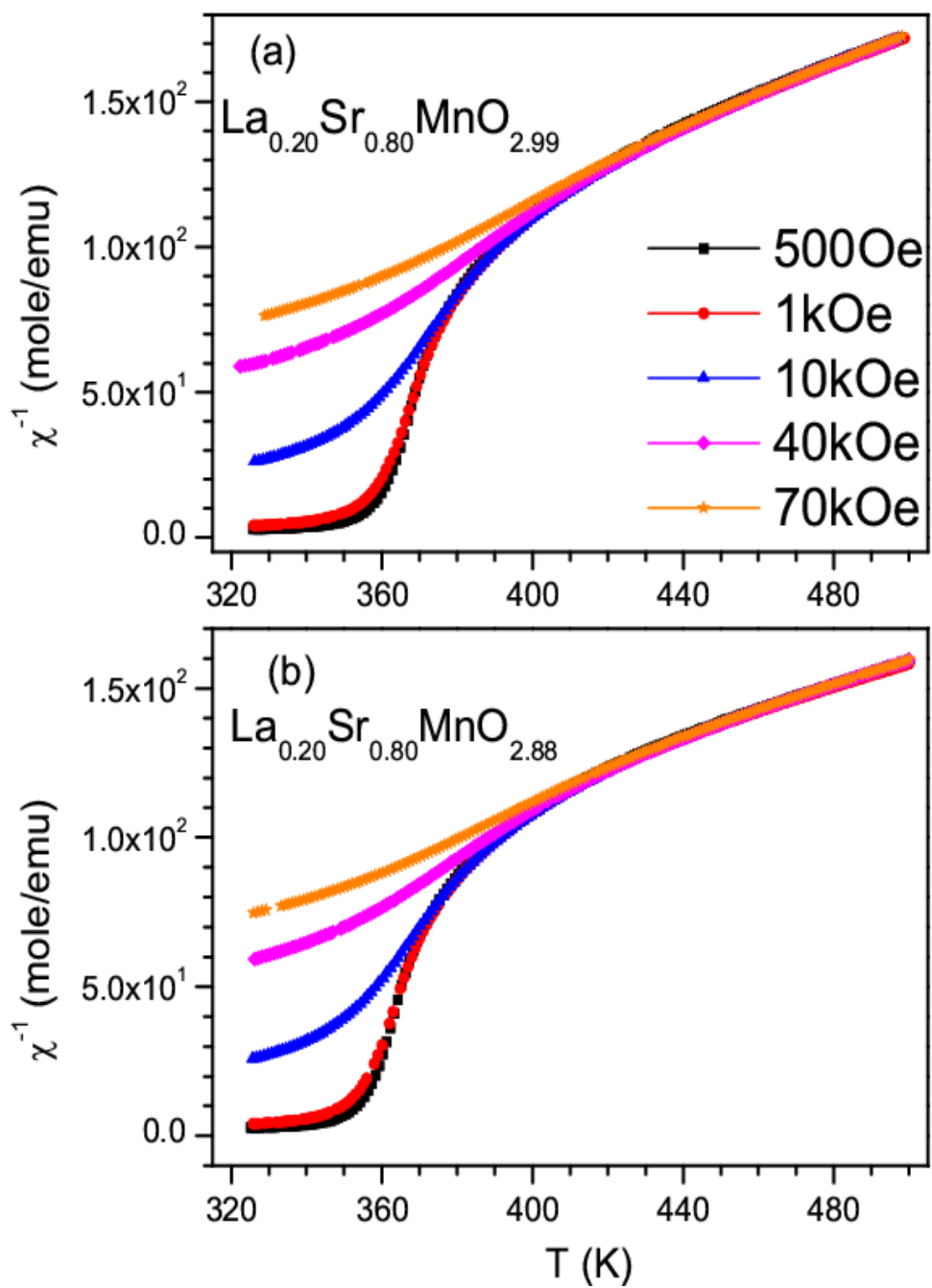

FIG 5.